\documentclass[reprint,showpacs,preprintnumbers,pre,superscriptaddress]{revtex4-1}
\usepackage{graphicx}
\usepackage[colorlinks=fals,linkcolor=blue]{hyperref}
\def\be{\begin{equation}}
\def\ee{\end{equation}}
\def\bea{\begin{eqnarray}}
\def\eea{\end{eqnarray}}
\begin{document}
\title{Structure--property relationships of cell clusters in biotissues: 2D analysis}
\author{Xiaohua Zhou}
 \affiliation{Department of Applied
Physics, Xi'an Jiaotong University, Xi'an 710049, People's Republic
of China}\affiliation{Department of Applied Statistics and Science, Xijing University, Xi'an 710123, China.}
\author{Erhu Zhang }
 \affiliation{Department of Applied
Physics, Xi'an Jiaotong University, Xi'an 710049, People's Republic
of China}
\author{Minggang Xia}
 \affiliation{Department of Applied
Physics, Xi'an Jiaotong University, Xi'an 710049, People's Republic
of China}
\author{Jianlin Liu}
 \affiliation{Department of Engineering Mechanics, China University of Petroleum, Qingdao 266555, China. }
\author{Shengli Zhang}\email{zhangsl@mail.xjtu.edu.cn}
 \affiliation{Department of Applied
Physics, Xi'an Jiaotong University, Xi'an 710049, People's Republic
of China}
\date{\today}
\begin{abstract}
To insight the relationships between the self-organizing structures of cells, such as the cell clusters, and the properties of biotissues is helpful in revealing the function and designing biomaterial. Traditional random foam model neglects several important details of the frameworks of cell clusters, in this study we use a more complete model, cell adhesion model, to investigate
the mechanical and morphological properties of the two-dimensional (2D) dry foams composed by cells. Supposing
these structures are formed due to adhesion between cells, the equilibrium formations
result from the minimum of the free energy. The equilibrium shape equations
 for high symmetrical structures without the volume constraint are derived, and the analytical results of the
corresponding mechanical parameters, such as the Young's modulus, bulk modulus
and failure strength, are obtained. Numerical simulation method is applied to study the complex shapes with the volume constraint and several stable multicellular structures are obtained. Symmetry-breaking due to the volume change is founded and typical periodic shapes and the corresponding phase transformations are explored. Our study provides a potential method to connect
the microstructure with the macro-mechanical parameters of biotissues.
The results also are helpful to understand the physical mechanism of how the structures of biotissues are formed.
\textbf{keywords:} {Soft stacking, Adhesion, Symmetry}
\end{abstract}

 \maketitle
\section{Introduction}
Soft stacking system which is composed by cells \cite{Skalak1981}, particles \cite{Seth} as well as other soft units \cite{Buchcic, Petroff, Pennybacker} often presents particular arrangement.
Comparing the soft stacking system with the atom stacking system in solid state physics, the former has more complex structures. A reason for inducing this complexity is that, unlike the rigid spherical atoms, the basic units in soft stacking system have strong adaptability to change their shapes under different circumstances. For example, cells can change their shapes to explore their environment \cite{Taloni}. Another reason is that, the composing unites in a soft stacking system possibly are multiplicate. Which implies that, when plenty of cells with different size (volume and surface area) are adhered together, the system will be very intricate. Due to the complexity, to insight the relationship between the configuration of soft stacking systems and the corresponding bio-functions still has challenge.

Adhesion which plays an important rule in forming the cell or vesicle self-organizing structures mainly derives from the interface interaction \cite{Engler, Schwarz}, protein interaction\cite{ZhangX} as well as charge interaction \cite{Pantazatos}. At molecular level to simulate the three kinds of interactions in a system is constrained by the complexity and long time expending. So, continuous model provides a valuable
option to reveal how the cell stacking system are formed due to adhesion. When the membrane proteins are ignored, cell membrane which is composed mostly by two layers of phospholipids can be taken as liquid bilayer vesicles as well as 2D surfaces. The equilibrium shapes for free vesicles are governed by the Helfrich-Canham bending energy theory \cite{Helfrich, Canham, Ouyang1}. Seifert and Lipowsky \cite{Seifert1,Lipowsky,Seifert2} extended this theory to deal with the vesicle adhering system. They derived the equilibrium shape equations and boundary conditions by minimizing the free energy and also conveyed that equilibrium shapes result from the competition between the elastic energy and the adhesion energy. In past two decades, their theory has been generally used to investigate vesicle adhesion configurations and great progress has been achieved. It explained the adhesion shape composed by several red blood cells \cite{Ziherl}. Deserno et al. \cite{Deserno} developed a general geometrical framework to derive the equilibrium shape equations and boundary conditions, which makes it possible to reveal the complex structures composed by a large number of cells.

\begin{figure} 
\includegraphics[scale =0.55]{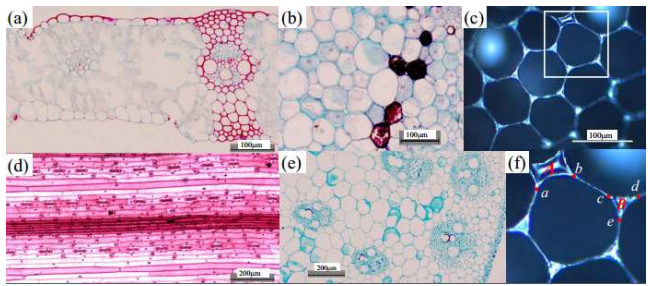}
\caption{ The adhesive stacking system of plant cells. (a) The cross section of wheat leaf.
We can see each epidermic cell adheres to its two neighbors. The dark cells adhered together are the vascular bundles. (b) The cross section of cotton tree. (c) The cross section of fresh-cut garlic sprout.
(d) The machine direction of wheat leaf in which the length of epidermic cell can reach about 1 mm.
(e) The cross section of cornstalk. (f) The partial enlarged view of the square region in (c).}
\label{fig1}
\end{figure}

Cells in plant tissues often present regular frameworks \cite{Villalon,Chen}. Some example are shown in Fig.~\ref{fig1}. We can see the length of epidermal cell can reach about 1 mm, but the width is no more than 40 $\rm{\mu}m$. Actually, the length of some plant cells can attain dozens of centimeters length. For example, the length of ramie cell can reach 145 mm, which is 3000 times of its width \cite{Pandey}. So to study the equilibrium shapes
of these tubular formations can be reduced to investigate
the shapes of their cross sections in 2D case. Also, these structures be previously named 2D dry foams and the properties have been wildly investigated \cite{Weaire1,Weaire2,Schliecker}. However, in these works several important details of the clusters of cells are neglected. In the solid foams, the walls of all cavities form a network and the joint points between them are rigid connections \cite{Schliecker,Roberts,Kraynik1,Vecchio}, that is different to the corresponding points in Fig.~\ref{fig1}(f). For example, the points $a$ to $e$ are the initial contact points between neighbor cells. Nearby each point, two edges of two neighbor cells have the asymptotic behavior to get close smoothly. After that, two edges will contact together until the next initial contact point. When the structure is under the external load, these initial contact points can shift along the edges and the contact length will change meanwhile. Further, in most cases, we can see in Fig.~\ref{fig1}(c) that three neighbor cells fence a triangular space, such as the area $B$. Also, four neighbor cells can fence a quadrangular space, such as the area $A$. These spaces do not belong to any cells. Each cell in bioissues has its own edges and volume. In the contact region, the edge is made up of two layers of individual cell edges. We can also pick up a cell from biotissues without breaking any other cells. These phenomena indicate the biotissues are formed by plentiful cells stacked together. But in most of solid foam models, the edges of cells are taken as an integrated network and the unique character of each cell is neglected.

Based on the above analysis, it needs a more complete model to study the frameworks of cells. In this model, the biotissue should be built up by cells and its macroscopical characters should rely on the mechanical properties of every single cell. We think the cell adhesion theory is a more suitable option. There are at lest two reasons. First, for single cell, we have known that the shape is determined by the minimization of the Helfrich-Canham bending energy. Particularly, although the plant cells have thick cell walls, we can take them as thin elastic shells and this theory also is suitable (neglect the in-plane strain) \cite{Tu}. Seconde, for plentiful cells, the cell adhesion theory not only considers the independence of singer cells but also contains the adhesion action between cells.
In this work, we use the cell adhesion theory to study the mechanical and morphological properties of the cell clusters in 2D case. In the
Sect.~II, the basic theory is shown and the numerical simulation method as well as the software are introduced. In the
Sect.~III, some analytical results for periodic structures without volume constraint are derived. In the
Sect.~IV, we show some numerical results obtained by simulation. Finally, these results are recapped in a short
discussion in Sec.~V.

\section{Theoretical methods }\label{model}
\subsection{The basic model}
In 2D case, let us consider a system which is composed by $n$ cells adhered together.
Supposing that the $i$th cell has $m$ neighbors adhered to it, its free energy is \cite{Seifert1,Lipowsky,Seifert2,Deserno}
\bea\label{e1}
E_i=\frac{1}{2}\kappa_i\oint \Lambda_i^2 ds_i -\frac{1}{2}\sum_{j=1}^{m} \omega_{ij} B_{ij}.
\eea
Here, $\kappa_i$ is the bending rigidity,
$\Lambda_i$ is the curvature, $ds_i$ is the element of the arc length of the cell, $\omega_{ij}$ and $B_{ij}$ are the adhesion potential (work of adhesion) and adhesion length between the $i$th cell and its $j$th neighbor, respectively. Considering the length and volume constraints, the corresponding energy functional is
\bea\label{e2}
\Omega_i=E_i+ \gamma_i\oint ds_i+ \Delta p_i\int\int d\sigma_i,
\eea
where $\gamma_i$ is the line tension coefficient, $\Delta p_i$ is the osmotic pressure difference between the inside and outside of the cell,
$d\sigma_i$ is the element of the area enveloped by the cell. For the whole system, the total energy is $E_t=\sum_{i=1}^{n}E_i$ and the energy functional is $\Omega_t=\sum_{i=1}^{n}\Omega_i$.
The equilibrium shapes are determined by the minimization of $\Omega_t$.

Supposing the whole system is in the $x-y$ plane. For the $i$th cell, let $\phi_i$
be the angles between the tangent of each arc and the $x$ axis, and define the clockwise direction as the positive direction of $\phi_i$, there is $\Lambda_i=d\phi_i/ds_i=\dot{\phi}_i$.
The equilibrium shape for whole system needs the first variation $\delta\Omega_t=0$. Due to each cell has
independent length $L=\oint ds_i$ and volume $V_i=\int\int d\sigma_i$, $\delta\Omega_t=0$ yields $\delta\Omega_i=0$ which gives
the general shape equation \cite{Ouyang1,ZhangS1,ZhangS2}
\bea\label{EqG}
\kappa_i\dddot{\phi}_i+\frac{1}{2}\dot{\phi}_i^3-\gamma_i\dot{\phi}_i+\Delta p_i=0.
\eea
Further, at the initial adhesion points between $i$th cell and $j$th cell, $\delta\Omega_t=0$ yields the boundary conditions \cite{Deserno}
\bea
\label{OB1}\kappa_i \dot{\phi}_i^2+\kappa_j \dot{\phi}_j^2-(\kappa_i+\kappa_j)\dot{\phi}_{ij}^2=2\omega,\\
\label{OB2}\frac{d}{ds}\Big[\kappa_i \dot{\phi_i}+\kappa_j \dot{\phi_j}-(\kappa_i+\kappa_j)\dot{\phi}_{ij}\Big]=0,\\
\label{OB3}\kappa_i \dot{\phi}_i+\kappa_j \dot{\phi}_j-(\kappa_i+\kappa_j)\dot{\phi}_{ij}=0.
\eea
Considering $\phi_i$ and $\phi_j$ turn clockwise in the $i$th cell and $j$th cell, respectively. In the adhesion region we define $\phi_{ij}$ belongs to the $i$th cell ($\phi_{ij}$ turns clockwise in the $i$th cell) and $\phi_{ji}$ belongs to the $j$th cell ($\phi_{ji}$ turns clockwise in the $j$th cell) and we have $\dot{\phi}_{ij}=-\dot{\phi}_{ji}$.

There is another expression for the Eq.~\ref{EqG} by using the Lagrange equation. Considering $\dot{\phi}_i=d\phi_i/ds_i$, there are
\bea\label{xy}
\dot{x}_i=\cos\phi_i,~~\dot{y}_i=\sin\phi_i.
\eea
Making use of $dV_i=x_i\sin\phi_i ds_i$, Eq.~\ref{e2} is reduced to
\bea\label{e3}
\Omega_i= \oint\Big[\frac{1}{2}\kappa_i \dot{\phi}_i^2 +\Delta p_i x_i\sin\phi_i+\gamma_i\Big] ds_i -\frac{1}{2}\sum_{j=1}^{m} \omega_{ij} B_{ij}.
\eea
Consequently, the Lagrange density can be written as \cite{Seifert1,Lipowsky,Seifert2}
\bea 
\nonumber\Pi_i(\phi_i,\dot{\phi}_i,x_i,\dot{x}_i,\lambda_i)=\frac{1}{2}\kappa_i \dot{\phi}_i^2 +\Delta p_i x_i\sin\phi_i+\gamma_i\\
+\lambda_i(\dot{x}_i-\cos\phi_i)-\frac{1}{2}\sum_{j=1}^{m} \omega_{ij},
\eea
where $\lambda_i$ is the Lagrange coefficient. The Lagrange equations $\frac{\partial\Pi_i}{\partial\phi_i}-\frac{d}{ds}\big(\frac{\partial\Pi_i}{\partial\dot{\phi}_i}\big)=0$ and $\frac{\partial\Pi_i}{\partial x_i}-\frac{d}{ds}\big(\frac{\partial\Pi_i}{\partial\dot{x}_i}\big)=0$ yield
\bea
\label{Eq01} \kappa_i\ddot{\phi}_i-\Delta p_i x_i
\cos\phi_i-\lambda_i\sin\phi_i=0,\\
\label{Eq02} \Delta p_i\sin\phi_i-\dot{\lambda}_i=0.
\eea
The Hamiltonian function is $ H=-\Pi_i+\dot{\phi}_i\frac{\partial\Pi_i}{\partial\dot{\phi}_i}+\dot{x}_i\frac{\partial\Pi_i}{\partial\dot{x}_i}$.
It has been proved that $H\equiv0$ \cite{Julicher} and it gives
\bea
\label{EqH}
\frac{1}{2}\kappa_i \dot{\phi}_i^2-\Delta p_i x_i\sin\phi_i+\lambda_i\cos\phi_i-\gamma_i=0.
\eea
It also has proved that Eqs.~\ref{xy}, \ref{Eq01}, \ref{Eq02} and \ref{EqH} are identical to the general shape equation \ref{EqG} \cite{Julicher}. Note that $\gamma_i$ only appears in Eq.~\ref{EqH}, so we can ignore Eq.~\ref{EqH} and only solve the left three equations because it can always be satisfied by choosing a suitable $\gamma_i$. This method will be used in following text.

In a multicellular system without any symmetry, whether the general shape equation \ref{EqG} or the Lagrange equations \ref{xy}, \ref{Eq01} and \ref{Eq02} are difficult to be solved analytically. In our work we suppose the
system is composed by equal cells which have the same physical parameters: $\kappa\equiv\kappa_i$,
$\omega\equiv\omega_{ij}$, $V\equiv V_i$ and the cell length $L\equiv\oint ds_i$. We will try to find analytical solutions with high symmetry for these Lagrange equations and use the finite element simulation method to study the complicate system and do stability analysis.

\subsection{Analytical method without volume constraint}

\begin{figure} 
\includegraphics[scale =0.55]{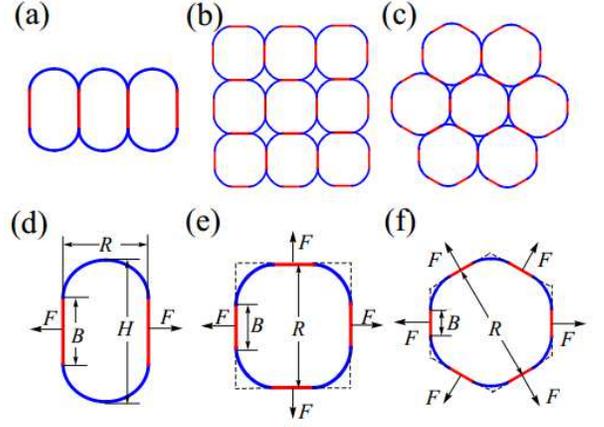}
\caption{ Three typical cell stacking modes, the
adhesion regions are in red color. (a) The 1D periodic adhesion; (b)
The foursquare stacking; (c) The hexagonal stacking. The (d), (e) and (f) are the
corresponding unit cell for each structure. The centre-to-centre spacing between two neighbor cells is $R$. The $F$ is the force acting on each cell along the periodic directions. The dot lines show the circumscribed polygons for the foursquare and hexagonal structures. Apparently, there are spaces between cells like the structure in Fig.~\ref{fig1}.}
\label{fig2}
\end{figure}

\begin{figure} 
\includegraphics[scale =0.55]{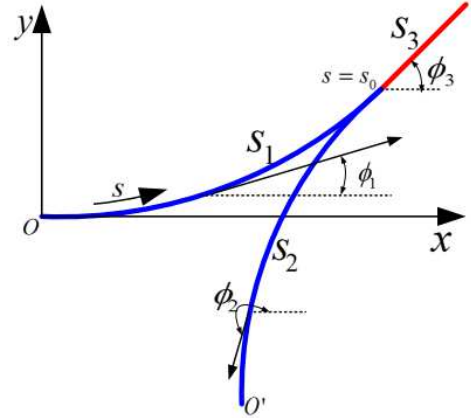}
\caption{ The basic unit for periodic structures. The $O$ and $O'$ points are the middle points of two free parts. The $S_1$ part begins at point $O$ and the point $O'$ is the end of $S_2$ part. These two parts initially contact at $s=s_0$. The $S_3$ part is half of the one adhesion part in Fig.~\ref{fig2}. For $S_1$ and $S_3$ parts, the arc length turns
anticlockwise. For $S_2$ part,
the arc length turns from point $s=s_0$ to $O'$. $\phi_i$ ($i=1,2,3$) is the angle between the tangent of each part and the $x$ direction.}
\label{fig3}
\end{figure}
When cells adhere together and form periodic
structures, there are many kinds of possible forms. Three typical forms are shown in Fig.~\ref{fig2}.
For simplicity, in each form we think the cells are equal and there is no volume constraint for each cell. In this case, we can make an ansatz that
all of the adhesion parts are straight lines. (This assumption will be confirmed by finite
element calculation in latter text). In these periodic structures, we can find the basic
unit as shown in Fig.~\ref{fig3}, which can be used to describe their equilibrium shapes.

In Fig.~\ref{fig3}, due to the symmetry, $S_1$ and $S_2$ parts are equal, thus we can only choose the $S_1$ and $S_3$ to study. The
Lagrange density in the region of $0\leq s\leq s_0$ for $S_1$ part is
\bea
 \label{Density1}&&\Pi_1 =\frac{1}{2}\kappa_1\dot{\phi}_1^2
    +\Delta p_1x_1\sin\phi_1+\lambda_1(\dot{x}_1-\cos\phi_1).
\eea
In the adhesion region $s_0\leq s\leq s_0+B/2$ , it is
\bea\label{Density2}
 \Pi_3 =\frac{1}{2}\kappa_3\dot{\phi}_3^2
    +\Delta p_3x_3\sin\phi_3+\lambda_3(\dot{x}_3-\cos\phi_3)-\omega,
\eea
where $\kappa_3=\kappa_1+\kappa_2$.
 The shape equations for the
$S_1$  part in $0\leq s\leq s_0$ can be written as
\bea
  \kappa_1\ddot{\phi}_1-\Delta p_1 x_1
\cos\phi_1-\lambda_1\sin\phi_1=0,\\
  \Delta p_1\sin\phi_1-\dot{\lambda}_1=0.
\eea
For the $S_3$ part in $s_0\leq s\leq s_0+B/2$, there are
\bea
  \kappa_3\ddot{\phi}_3-\Delta p_3 x_3
\cos\phi_3-\lambda_3\sin\phi_3=0,\\
  \Delta p_3\sin\phi_3-\dot{\lambda}_3=0.
\eea
At the initial contact point $s=s_0$, the boundary conditions in Eqs.~\ref{OB1}, \ref{OB2} and \ref{OB3}
are changed to
\bea
\label{contact1}\kappa_1\dot{\phi}_1^2+\kappa_2\dot{\phi}_2^2-\kappa_3\dot{\phi}_3^2=2\omega,\\
\label{contact2}\kappa_1\ddot{\phi}_1+\kappa_2\ddot{\phi}_2-\kappa_3\ddot{\phi}_3=0,\\
\label{contact3}\kappa_1\dot{\phi}_1+\kappa_2\dot{\phi}_2-\kappa_3\dot{\phi}_3=0.
\eea
Considering that all cells are equal, we have $\kappa_1=\kappa_2=\kappa_3/2=\kappa$.
Note that the $S_1$ and
$S_2$ parts should be equal in a periodic system, and that the rotation directions
of them are opposite in the adhesion region, at the point $s=s_0$, there are
\bea\label{self-consistent}
\dot{\phi}_1=-\dot{\phi}_2,~~\ddot{\phi}_1=-\ddot{\phi}_2,~~\dot{\phi}_3=\ddot{\phi}_3=0.
\eea
The above conditions make sure that Eqs.~\ref{contact2} and
\ref{contact3} can be satisfied, and Eq.~\ref{contact1} is reduced to
\bea\label{W1}
\omega=\kappa\dot{\phi}_1^2,~(s=s_0).
\eea
 Due to each adhesion part is a straight line, $\phi_3$ should be independent to $s$ and we choose $\phi_3\equiv\phi_0=\pi/n$. One can find that,
$n=2,~4$ and $6$ are corresponding
to the shapes in Fig.~\ref{fig2}(d), (e) and (f), respectively. In order to obtain dimensionless results, we choose $\kappa=1$ and fix the girth of each cell $\oint ds = 2\pi$ \cite{Zhou1}.

First, we study the structures without volume constraint for each cell, which means the pressure difference $\Delta p_i=0$. (The non zero pressure
case will be discussed in latter text). For convenient, in the following text we choose $\lambda_1=\lambda$, $\phi_1=\phi$, and $x_1=x$. The shape equations for the $S_1$ part can be reduced to
\bea
\label{shape1} \ddot{\phi}=\lambda\sin\phi,~(0<s<s_0).
\eea
The first integral is
\bea
\label{shape2} \dot{\phi}^2=2\lambda(C-\cos\phi),
\eea
where $C$ is an integral constant. Consequently, we have
\bea
\label{shape3} ds=\frac{d\phi}{\sqrt{2\lambda(C-\cos\phi)}}.
\eea
Choosing the initial conditions $x(0)=0$, $y(0)=0$ and $\phi(0)=0$ and
using the Eq.~\ref{xy}, the coordinates at the point $s=s_0$ are
\bea
\label{coordinates-X} \nonumber
  x_0 &=&\int_0^{s_0}\cos\phi ds =\int_0^{\phi_0}\frac{\cos\phi d\phi}{\sqrt{2\lambda(C-\cos\phi)}}\\\nonumber
       &=&\sqrt{\frac{2}{\lambda(C-1)}}\times\Bigg[C\times\text{FN}[\frac{\phi_0}{2},\frac{2}{1-C}]\\
       & &-(C-1)\times\text{SN}[\frac{\phi_0}{2}, \frac{2}{1-C}]\Bigg],\\
\label{coordinates-Y}\nonumber
  y_0  &=& \int_0^{s_0}\sin\phi ds=\int_0^{\phi_0}\frac{\sin\phi d\phi}{\sqrt{2\lambda(C-\cos\phi)}}\\
       &=&\sqrt{\frac{2}{\lambda}}\big(\sqrt{C-\cos\phi_0}-\sqrt{C-1}\big),
\eea
where $\text{FN}[x,y]$ and $\text{SN}[x,y]$ are the first and second
incomplete elliptical integral, respectively.
The length constraint $L=\oint ds=2\pi$ gives the following equation
\bea
\label{L1}s_0+\frac{B}{2}=\int_0^{\phi_0}\frac{d\phi}{\sqrt{2\lambda(C-\cos\phi)}}+\frac{B}{2}=\phi_0.
\eea
It yields
\bea
\label{L2}\sqrt{\frac{2}{\lambda(C-1)}}\times\text{FN}[\frac{\phi_0}{2},
\frac{2}{1-C}]+\frac{B}{2}=\phi_0.
\eea
Making use of Eq.~\ref{shape2}, condition \ref{W1} is
changed as
\bea
\label{W2}\lambda=\frac{\omega}{2(C-\cos\phi_0)}.
\eea
Substituting the above equation into Eq.~\ref{L2}, we get
\bea
\label{B}B=2\phi_0-4\sqrt{\frac{C-\cos\phi_0}{\omega(C-1)}}\times\text{FN}[\frac{\phi_0}{2},
\frac{2}{1-C}].
\eea
The diameters
of the inscribed circle and circumscribed circle of each cell in Fig.~\ref{fig2} are
\bea
\label{R}R&=&2x_0\csc\phi_0+B\cot\phi_0,\\
\label{H}H&=&2x_0\cot\phi_0+B\csc\phi_0+2y_0.
\eea

 The dimensionless total energy for each cell is
\bea\label{Ed1}
\nonumber E&=&\frac{2\pi}{\phi_0}\bigg(\int_0^{S_0} \frac{1}{2}\dot{\phi}^2dS-
\frac{1}{4}\omega B\bigg)\\
\nonumber &=&\frac{2\pi}{\phi_0}\int_0^{\phi_0}
\frac{\lambda(C-\cos\phi)}{\sqrt{2\lambda(C-\cos\phi)}}d\phi-
\frac{\pi}{2\phi_0}\omega B\\
&=&\frac{2\pi}{\phi_0}\sqrt{2\lambda(C-1)}\times\text{SN}[\frac{\phi_0}{2},
\frac{2}{1-C}]-\frac{\pi\omega B}{2\phi_0}.
\eea
Substituting Eqs.~\ref{W2} and \ref{B} into the above equation, we
obtain
\bea\label{Ed2}
  \nonumber
  E&=&\frac{2\pi}{\phi_0}\sqrt{\frac{\omega(C-1)}{C-\cos\phi_0}}\times\text{SN}[\frac{\phi_0}{2},\frac{2}{1-C}]-\pi \omega \\
   & &+\frac{2\pi}{\phi_0}\sqrt{\frac{\omega(C-\cos\phi_0)}{C-1}}\times\text{FN}[\frac{\phi_0}{2},\frac{2}{1-C}].
\eea
In the above equation, if $\omega$ and $\phi_0$ are known, there is only an unknown constant $C$. Defining $\chi=1/C$, the equilibrium shapes satisfy
\bea\label{equalibrium}
\frac{dE}{dR}=\frac{dE/d\chi}{dR/d\chi}=0.
\eea
We find this equation yields $\chi\rightarrow0$ ($C\rightarrow\pm\infty$).
Then substituting Eq.~\ref{W2} into Eq.~\ref{shape2}, we have
\bea
\dot{\phi}^2=\frac{\omega (C-\cos\phi)}{C-\cos\phi_0}=\omega.
\eea
This result indicates that the optimal shapes of the free parts in
Fig.~\ref{fig2} are circular arcs
 with the same radius
$r_0=\sqrt{1/\omega}$ and the total energy for each cell
$E=2\pi\sqrt{\omega}-\pi \omega$. Then the equilibrium distance between two cell is
\bea\label{R0}
\nonumber R_0&=&2r_0+B\cot\phi_0\\
&=&2\sqrt{1/\omega}+2(1-\sqrt{1/\omega})\phi_0\cot\phi_0.
\eea
Note that the total length of each cell is fixed to $2\pi$, it needs the equilibrium shapes satisfy $r_0<1$. It yields $\omega>1$, that is the basic condition to form the cell-cell adhesion structures. If the adhesion occurs between a cell and a rigid plane, it needs $\omega>0.5$ \cite{Seifert1,Zhou2}.

\subsection{Simulation method and the software}

To study the complicate adhesion system, we use the software \emph{Surface Evolver} \cite{Brakke} which has been generally applied to simulate the equilibrium shapes of single vesicle \cite{Yan,Zhou3}, cell adhesion system \cite{Ziherl}, random foam \cite{Kraynik2} and droplet adhesion structure \cite{Lv}. First, it needs to build an initial geometric model for the structure. Here we give an example model for three cells' adhered together in the \emph{Supplementary Material}. Next, we need to define the energy functional of the system. For one cell it is
\bea\label{m}
\nonumber \Omega_i=\alpha \oint \Lambda_i^2 ds_i -\frac{1}{2}\sum_{j=1}^{m} \omega_{ij} B_{ij}\\
+ \gamma_i\oint ds_i+ \Delta p_i\int\int d\sigma_i.
\eea
where $\alpha$ is a constraint and we set $\alpha=1/2$. Also, we fix $L=\oint ds_i=2\pi$. Defining the reduced volume $v_i=V_i/\pi=\int\int d\sigma_i/\pi$ we have $V_i=\int\int d\sigma_i=\pi v_i$ and $0<v_i\leq 1$ due to $L=2\pi$. Then we can fix the volume $V_i$ by setting the value of $v_i$. In this software the adhering potential $\omega_{ij}$ is named ``tension" of the edges. After setting these parameters: $\alpha, L, \upsilon_i$ and $\omega_i$, we can use the \emph{Surface Evolver} to find the equilibrium shape. In this process the perturbation method is used to find the optimal deformation direction of the system step by step and finally induces the shape converges to the equilibrium state with lower energy. More details of this can be obtained from Brakk's home page \cite{Brakke}. When it reaches the equilibrium state, the corresponding $B_{ij}$, $\gamma_i$, $\Delta p_i$ and the energy $E_i$ can also be obtained. The above method is used to find the solution of Eq.~\ref{EqG} numerically.

Although $\delta\Omega_t=0$ yields the equilibrium shapes equation Eq.~\ref{EqG}, we cannot make sure each solution of Eq.~\ref{EqG} is stable because stable shapes need $\delta^2\Omega_t>0$. In the \emph{Surface Evolver}, it can calculate the so called \emph{Hessian},
which is the matrix of the second order differential coefficient of the total energy.
At an equilibrium point, if the \emph{Hessian} is positively defined, this means the point is a strict local minimum. Here we can give a simple explain about this method. In the finite elemental method, a continuous curve is divided into many straight lines. Then the elastic energy functional can be written as $\Omega(x_i)$, where $x_i$ is the coordinates of the vertices of two neighbor straight lines. At $x_i=x_{i0}$, we expand the energy functional $\Omega(x_{i0}+\Delta x)=\Omega(x_{i0})+a_i \Delta x_i+b_i (\Delta x_i)^2+\cdots $, where $a_i=\frac{\partial \Omega}{\partial x_i}\mid_{x_i=x_{i0}}$ and $b_i=\frac{\partial^2\Omega}{\partial x_i^2}\mid_{x_i=x_{i0}}$. If the shape reaches an equilibrium state, there is $a_i=0$. Further, a stable shape also needs $b_i>0$. Acutely, $x_i$ should be a vector and $b_i$ should be a matrix which is named the \emph{Hessian} in the \emph{Surface Evolver}. More details of the \emph{Hessian} matrix can be obtained from the introduction file of this software \cite{Brakke}. After a longtime simulation, the shape will converge to an equilibrium state. Then using the \emph{Hessian} order we can obtain wether the \emph{Hessian} is positively defined or not. If it is, we can say the shape is stable. But if it is not, the shape possibly is not stable and needs a longtime simulation again.

\section{Analytical Results }\label{Analytical results}

\subsection{The periodic bead-like shape without volume constraint}
\begin{figure} 
\includegraphics[scale =0.55]{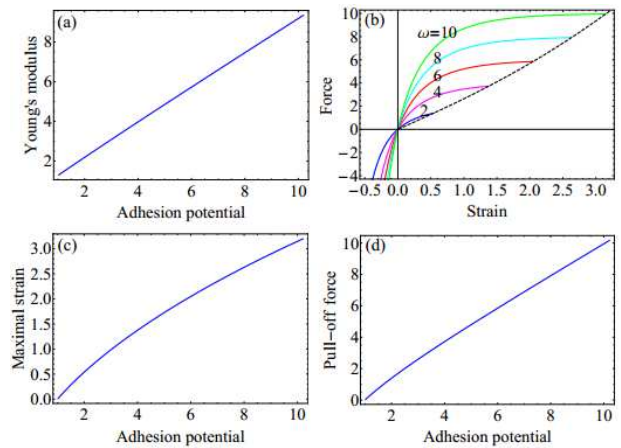} \caption{
Mechanical properties for the structure in Fig.~\ref{fig2}(d). (a) The Young's modulus for different adhesion potential $\omega$. (b) The force-strain relationships for different $\omega$. The dashed line is the adhesion failure boundary. (c) The maximal strain for different $\omega$. (d) The pull-off force for different $\omega$.}
\label{fig4}
\end{figure}

When $\phi_0=\pi/2$, we get the bead-like shape in Fig.~\ref{fig2}(a). For the optimal shape we define the elastic coefficient
\bea\label{K1}
K=\frac{d^2E}{dR^2}\bigg|_{R=R_0}=\frac{d^2E/d\chi^2}{(dR/d\chi)^2}\bigg|_{\chi\rightarrow0}.
\eea
The corresponding Young's modulus is
\bea\label{Y}
Y=KR_0/H_0,
\eea
where $H_0=H(\chi\rightarrow0)$. Fig.~\ref{fig4}(a) shows the relationship between Young's modulus and $\omega$, which indicates that $Y$ is nearly linear with $\omega$. Furthermore, we
can define the Poisson's ratio
\bea\label{P}
\mu=-\frac{dH}{dR}\bigg|_{R=R_0}=-\frac{dH/d\chi}{dR/d\chi}\bigg|_{\chi\rightarrow0}=0.6369.
\eea
It is a constant and we hope this results can be
tested by the future experiments.

Now we consider that the equilibrium bead-like structure is under the action of a couple of forces along the length directions and the structure is in balance. Fig.~\ref{fig2}(d) shows the diagram of one cell constrained by a couple of forces. Then the
force is
\bea\label{F1}
F=\frac{dE}{dR}=\frac{dE/d\chi}{dR/d\chi},
\eea
and the strain is
\bea\label{s1}
\xi=(R-R_0)/R_0.
\eea
Where $F=F(\chi)$ and $\xi=\xi(\chi)$ only depend on $\chi$. So we obtain the relationship between $F$ and $\xi$, which is shown in Fig.~\ref{fig4}(b) for different $\omega$. However, the adhesion structure will be failure when the adhesion length $B$ reduces to zero. Letting $B=0$ in Eq.~\ref{B}, we get $\chi=\chi_m$. Substituting it into Equation~\ref{s1}, we obtain the maximal strain $\xi_m=\xi(\chi_m)$. Fig.~\ref{fig4}(c) shows the relationship between $\xi_m$ and $\omega$. Also, we can get the pull-off force $F=F(\chi_m)$ for different $\omega$ as shown in Fig.~\ref{fig4}(d).

\subsection{The square shape and the hexagonal shape without volume constraint} \label{sec:4}
\begin{figure} 
\includegraphics[scale =0.55]{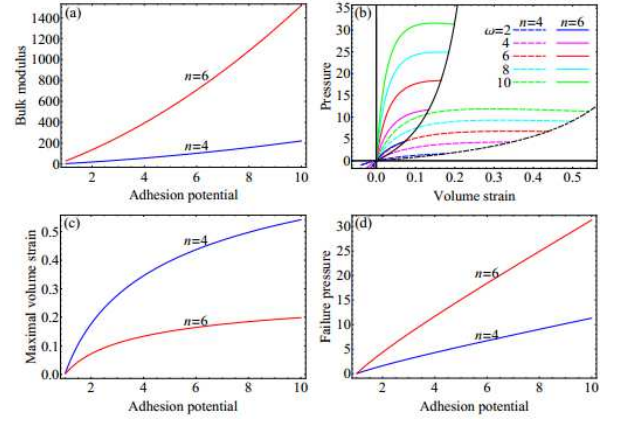} \caption{
Machanical properties for square structure (n=4) and hexagonal structure (n=6). (a) Relationship between bulk modulus and adhesion potential. (b) Relationship between pressure and volume strain. The two black lines are the adhesion failure boundaries. (c) The maximal strain for different adhesion potentials. (d) The failure pressure (failure strength) for different adhesion potential.}
\label{fig5}
\end{figure}

When $\phi_0=\pi/4$, we obtain the square shape in Fig.~\ref{fig2}(e). For the optimal structure, we define the elastic coefficient
\bea\label{KS}
K_S=\frac{1}{2}\frac{d^2E}{dR^2}\bigg|_{R=R_0}=\frac{1}{2}\frac{d^2E/d\chi^2}{(dR/d\chi)^2}\bigg|_{\chi\rightarrow0}.
\eea
If there are small perturbations $\Delta R$ along the square edges, the corresponding force is $\Delta F=K_S \Delta R$. The pressure on each edge of the square is $\Delta P=\Delta F/R_0=K_S\Delta R/R_0$. The valid area (the area for the circumscribed square) for the cell is $A_0=R_0^2$ and the area perturbation $\Delta A=2R_0\Delta R$. So, the bulk modulus (actually the area modulus in 2D case) is
\bea\label{KA4}
K_A=-\frac{dP}{dA}A_0=\frac{\Delta P}{\Delta A}A_0=\frac{1}{2}K_S.
\eea
When $\phi_0=\pi/6$, using the similar method, we obtain the elastic coefficient as
\bea\label{KH}
K_H=\frac{1}{3}\frac{d^2E}{dR^2}\bigg|_{R=R_0}=\frac{1}{3}\frac{d^2E/d\chi^2}{(dR/d\chi)^2}\bigg|_{\chi\rightarrow0}.
\eea
The bulk modulus for the optimal shape in Fig.~\ref{fig2}(f) is
\bea\label{KA6}
K_A=\frac{\sqrt{3}}{2}K_H.
\eea
We show the bulk modulus in Fig.~\ref{fig5}(a). Apparently, with the similar $\omega$, the hexagonal
structure has a bigger bulk modulus than the square structure.

As shown in Fig.~\ref{fig2}(e), two couples of forces are supposed to act on this square structure and keep it in balance.
The pressure on each edge of the circumscribed square derived from the out side forces is
\bea\label{P4}
P=\frac{1}{2R_0}\frac{dE}{dR}=\frac{1}{2R_0}\frac{dE/d\chi}{dR/d\chi}.
\eea
Correspondingly, for the hexagonal structure in Fig.~\ref{fig2}(f), the pressure on each edges of the circumscribed hexagon is
\bea\label{P6}
P=\frac{\sqrt{3}}{3R_0}\frac{dE}{dR}=\frac{\sqrt{3}}{3R_0}\frac{dE/d\chi}{dR/d\chi}.
\eea
The volume strain for the two kinds of shapes induced by the force is $\sigma=R^2/R_0^2-1$.
Fig.~\ref{fig5}(b) shows the relationship between pressure $P$ and the strain $\sigma$ for the square structure (dashed lines) and
the hexagonal structure (solid lines). Let $B=0$, we obtain the maximal strain shown in Fig.~\ref{fig5}(c), which yields
the adhesion failure boundaries (the two black lines) in Fig.~\ref{fig5}(b). Moreover, we show the maximal strain and failure
pressure for each structure under different adhesion potential in Fig.~\ref{fig5}(d), which can be taken as the failure strength for these structures.

\section{numerical results } \label{numerical results}
\subsection{The multicellular bead-like structures}
\begin{figure} 
\includegraphics[scale =0.55]{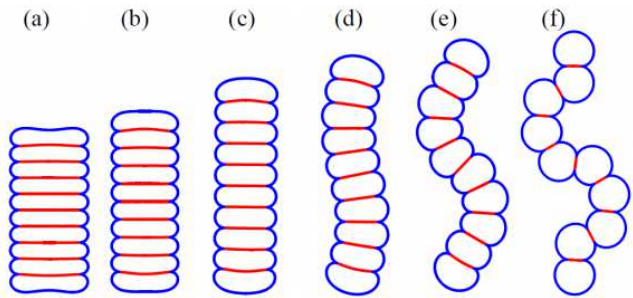}
\caption{Symmetry breaking for bead-like structure composed by 10 cells with $\omega=8$. From (a) to (f) $v=0.5,0.6,0.7,0.8,0.9$ and 0.98 respectively. The symmetry breaking occurs at $v\approx0.75$.}
\label{fig6}
\end{figure}

\begin{figure} 
\includegraphics[scale =0.55]{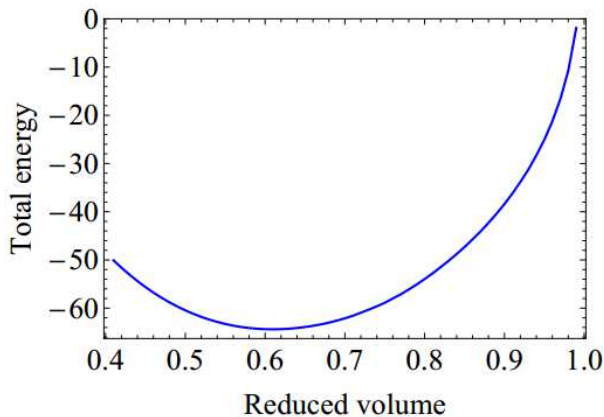}
\caption{The relationship between the total energy and the reduced volume for bead-like structure composed by 10 cells with $\omega=8$. At $v\approx0.6$, energy has the minimal value.}
\label{fig7}
\end{figure}

\begin{figure} 
\includegraphics[scale =0.55]{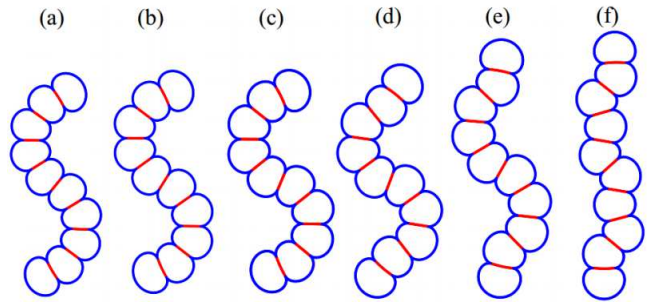}
\caption{ Deformation induced by the change of adhesion potential for bead-like structure composed by 10 cells with $v=0.94$. From (a) to (f) the $\omega=4,5,6,7,8$ and 9, respectively.}
\label{fig8}
\end{figure}

First, the multicellular bead-like structure are studied. In Fig.~\ref{fig6}, each structure is composed by 10 cells with $\omega=8$. Gradually changing $v$, the symmetry breaking is found. When $v<0.75$, the formation is a straight line and has $D_2$ symmetry. When $v>0.75$, the structure will bend and presents periodic formation. Following the increase of $v$, the amplitude of the structure will increase and the period length tends to decrease. It means the cell number contained in one period is negatively related to $v$. Similar symmetry breaking induced by changing adhesion potential ware found between two red blood cells \cite{Ziherl}. For fixed $\omega$, we find there is an optimal $v$, at which the energy reaches its minimum. As shown in Fig.~\ref{fig7}, the total energy
has the minimal value nearby $v=0.6$. A theoretical explain will be shown in latter text.

Fig.~\ref{fig8} shows the deformation induced by the change of adhesion potential at $v = 0.94$. We can see that, with the increase of $\omega$, the amplitude and the cell number contained in one period will decrease. These waved 2D structures in Fig.~\ref{fig6} and Fig.~\ref{fig8} imply that the bead-like 3D shapes possibly present curved formations as well as helical structures. Some kinds of bacteria and algaes present bead-like structures, such as streptococcus.

\subsection{The dense stacking systems}
\begin{figure} 
\includegraphics[scale =0.55]{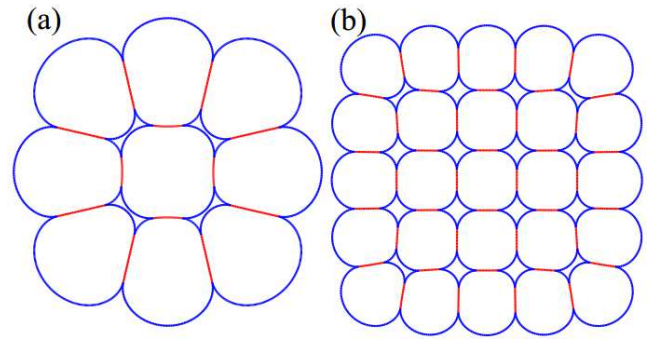}
\caption{ Two stable structures with $D_4$ symmetry at $\omega=4$. (a) Stable adhesion structure composed by 9 cells and $v=0.95$. (b) Stable adhesion structure composed by 25 cells and $v=0.97$.}
\label{fig9}
\end{figure}
Fig.~\ref{fig9} depicts two kinds of formations which are assumed to investigate the square adhesion systems in which each cell adheres to its four neighbors and the system can be extended to the whole 2D space. For Fig.~\ref{fig9}(a), the shape appears in the region $0.94<v<1$ and always keeps the $D_4$ symmetry. When $v<0.94$, the free parts of the central cell will adhere to the other four neighbor cells which do not adhere to it before. As to Fig.~\ref{fig9}(b), the stable region is $0.96<v<1$ and it also has the $D_4$ symmetry in this region. When $v<0.96$, the four cells in the corner of the out square will adhere to their nearest neighbors in the middle layer.

\begin{figure} 
\includegraphics[scale =0.55]{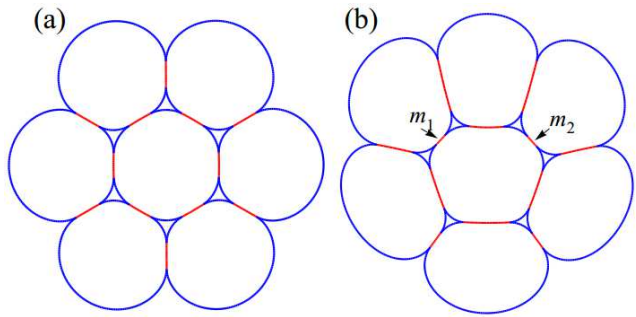}
\caption{ Stable adhesion structures composed by 7 cells with $\omega=4$. (a) $v=0.99$. (b) $v=0.96$.}
\label{fig10}
\end{figure}
\begin{figure} 
\includegraphics[scale =0.55]{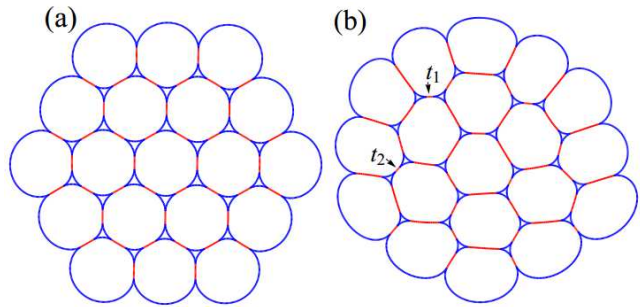}
\caption{ Stable adhesion structures composed by 19 cells with $\omega=4$. (a) $v=0.99$. (b) $v=0.96$.}
\label{fig11}
\end{figure}

Fig.~\ref{fig10} shows two kinds of structures composed by 7 cells. The shape in Fig.~\ref{fig10}(a) nearly has $D_6$ symmetry. Gradually decreasing $v$, the structure changes to the shape in Fig.~\ref{fig10}(b) at $v=0.96$ with $D_1$ symmetry. When $v<0.92$, the two outer cells in Fig.~\ref{fig10}(b) will lose the adhesion to the central cell due to the $m_1$ and $m_2$ adhesion parts reduce to zero. Fig.~\ref{fig11} shows two kinds of structures composed by 19 cells. The shape in Fig.~\ref{fig11}(a) nearly has $D_6$ symmetry. But following the decrease of $v$, the non-symmetrical configuration will become manifest and the shape will change to Fig.~\ref{fig11}(b). When $v<0.98$, our simulation indicates that there are many different shapes which have very similar total energy and the shape will change from one to one following the increase of simulation time. We think they are degenerate energy shapes and the energy barriers between them are very low. The degeneracy will increase with the decrease of $v$. When $v<0.92$, the two outer cells in Fig.~\ref{fig11}(b) will lose the adhesion to the middle layer cells due to the $t_1$ and $t_2$ adhesion parts reduce to zero.

Our simulation also reveals that, when old adhesion parts become separate, new adhesion parts will probably appear somewhere. Unfortunately, at present time it is very difficult to track the evolvement of adhesion because the adhering or departing process between two cells can not be achieved freely in simulation. One possible way is to construct every possible formation and simulate their evolution processes. So, we can obtain the whole evolutionary path by connecting each independent process. However, this method is extraordinarily time-consuming when a stacking system contains plentiful cells.

\subsection{Periodic Shapes}

In the former section, periodic shapes without volume constraint is studied analytically. If we consider the volume constraint, it is very difficult to obtain analytical results. Numerical simulation is a feasible way to find stable shapes. But at first, we tested our former analytical results with $\Delta p_i= 0$ by the \emph{Surface Evolver}. Our simulation indicates that, without volume constraint, all of the adhesion parts in the periodic structures in Fig.~\ref{fig2} are straight lines and our former analytical results with $\Delta p_i= 0$ are reliable.

\begin{figure} 
\includegraphics[scale =0.55]{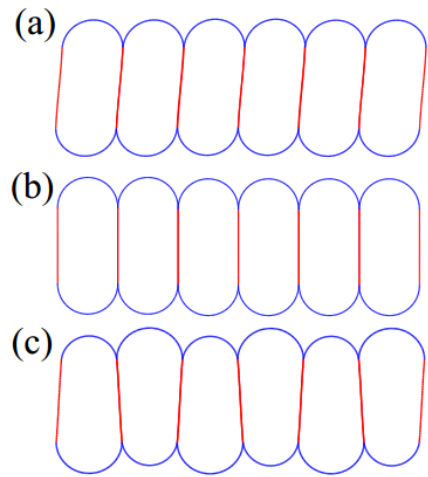}
\caption{ A phase transformation between the staircase structure and the zig-zag shape at $\omega=4$ and $v=0.8$. (a) A stable staircase structure which contains one cell for one period with $R=1.11$ and $\Delta p=2.48$. (b) A critical state in which each cell is nearly composed by two half of circles and two straight lines, which is the unstable saddle point $b$ in Fig.~\ref{fig13} with $R=1.1056$. (c) A stable zig-zag structure which contains two cells for one period with $R=1.10$ and $\Delta p=-2.65$.}
\label{fig12}
\end{figure}

\begin{figure} 
\includegraphics[scale =0.55]{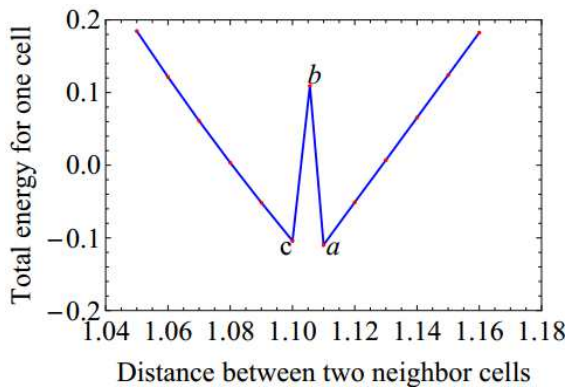}
\caption{ The energy-distance curve of one cell for periodic bead-like structure with $\omega=4$, $v=0.8$. There is an energy barrier nearby a critical state with $R=1.1056$. The a, b and c points are corresponding to the shapes in Fig.~\ref{fig12}(a), (b) and (c), respectively.}
\label{fig13}
\end{figure}

Further, three kinds of periodic structures with volume constraint for each cell are investigated.
In the \emph{Surface Evolver}, obtaining a stable periodic shape needs a long time searching.
Fig.~\ref{fig12}(a) shows a
stable bead-like structure, in which a period contains one cell. To attain this shape, we need to scan the centre-to-centre spacing $R$ between two neighbor periods (note that a period can contain two or more cells) and find out the optimal distance $R$ at which one period has the lowest total energy. An example is shown in Fig.~\ref{fig13}. Particularly, lots of simulations indicate that the optimal distance is close to the following analytic value. For a given $v$, supposing each cell is composed by two half of circles and two straight lines like the shape in Fig.~\ref{fig2}(d), the distance between two neighbor cells is
\bea\label{Rs}
R_s=2-2\sqrt{1-v},
\eea
and the total energy for each cell is
\bea\label{ET1}
E_t=\pi/v+\pi(1/v-\omega)\sqrt{1-v}.
\eea
For example, when $v=0.8$, we have $R_s=1.10$ and $E_t=0.06$, which are close to the simulation values in Fig.~\ref{fig13}. Besides the staircase shape in Fig.~\ref{fig12}(a), it also reveals that there is another stable bead-like shape as shown in Fig.~\ref{fig12}(c), in which a period contains two similar cells. Between these two stable shapes, it is the critical structure shown in Fig.~\ref{fig12}(b) that one cell is composed by two half of circles and two straight lines. The centre-to-centre spacing between two neighbor cells satisfies $R=R_s$. But this critical shape is unstable. Fig.~\ref{fig13} shows the energy-distance relation obtained by simulation, which indicates there is an energy barrier between the two stable structures. The critical shape is on the unstable saddle point and the energy change in phase transformation is discontinuous. Therefore, it is possibly a second-order phase transformation between the two stable shapes and can be induced by the change of distance. Simulation also indicates this phase transformation can be found in the whole volume region ($0<v<1$). Actually, similar results have been reported \cite{Ziherl} to explain the erythrocyte stacking structures.

\begin{figure} 
\includegraphics[scale =0.55]{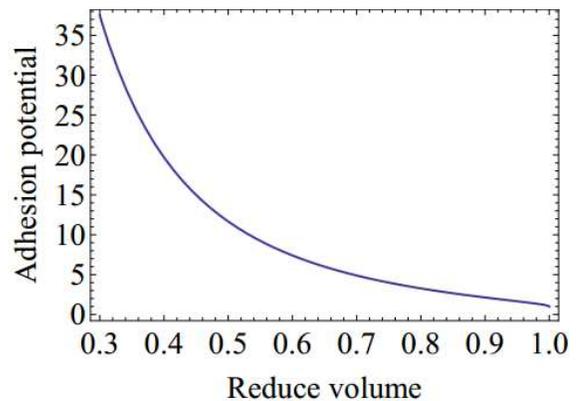}
\caption{ The curve of $\omega$ vs $v$ drives from Equation~\ref{w-v}.}
\label{fig14}
\end{figure}

If $\omega$ is fixed, the energy in Equation~\ref{ET1} has the minimum at a suitable volume. Let $dE_t/d v =0$, we get
\bea\label{w-v}
v=(2\sqrt{\omega}-1)/\omega.
\eea
Fig.~\ref{fig14} shows the curve of $\omega$ vs $v$. Our simulation also agrees with this theoretic result. For example, Fig.~\ref{fig3} has shown that the total energy for the 10 cells' bead-like structure has the minimal value nearby $v=0.6$ with $\omega=8$. Choosing $\omega=8$, Equation~\ref{w-v} gives $v\approx0.58$.

When comparing the periodic shapes in Fig.~\ref{fig12} and the multicellular adhesion structures in Fig.~\ref{fig6} and Fig.~\ref{fig8}, there are two evident differences.
 First, there are $D_2$ symmetrical structures for multicellular adhesion system, such as
 Fig.~\ref{fig4}(a), (b) and (c). But we didn't find stable $D_2$ symmetrical formations for periodic bead-like shapes. Second, the staircase configuration in Fig.~\ref{fig12}(a) only appeals in the periodic structures and there is no similar stable shapes for the multicellular shapes.

\begin{figure} 
\includegraphics[scale =0.55]{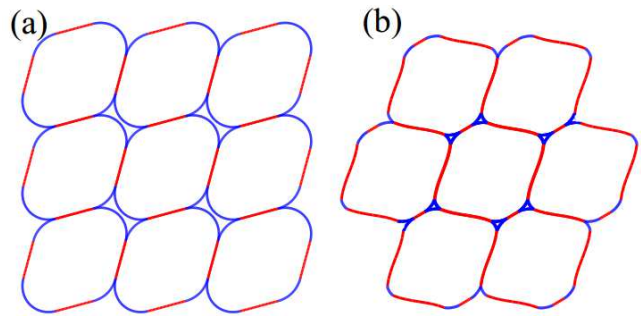}
\caption{ Two kinds of typical stable periodic structures. (a) Each cell contacts to its four neighbors and the structure has the equal period $R=1.774$ in two upright directions with $\omega=4$ and $v=0.94$.
(b) Each cell contacts to its six neighbors with $\omega=4$ and $v=0.90$. This structure has the equal period $R=1.77$ in two directions with the include angel $\pi/3$.}
\label{fig15}
\end{figure}

Very specially, if Equation~\ref{R0} is satisfied, the critical shape which is composed
by two half of circles and two straight lines will be an analytical solution. Then $R_0=R_s$ coincidentally yields the same equation to Equation~\ref{w-v}. Therefor, Equation~\ref{w-v} gives an analytical global optimal solution for periodic bead-like shapes. When Equation~\ref{w-v} is satisfied, our simulation reveals that the energy barrier is vanished and the critical shape is a stable middle phase between the staircase structure and the zig-zag structure. Therefor, the energy barrier only occurs when Equation~\ref{w-v} can not be satisfied.

Besides the bead-like shapes, two other periodic formations are investigated. Fig.~\ref{fig15}(a) shows a stable rhombus structure which has two periods on two upright directions. The Surface Evolver code for this model is shown in the \emph{Supplementary Material}. Our simulation indicates that the stable shape needs the periodic length on two upright directions are equal. It also reveals that the optimal distance is close to the following analytic value. For a given $v$, supposing that each free part of the rhombus shape is a quarter of circle with the same radii and the adhesion parts are equal straight lines, the distance between two neighbor cells is
\bea\label{Rs4}
R_s=2-2\sqrt{(1-v)(4-\pi)}.
\eea
It gives $R_s=1.773$ for $\omega=4$. Fig.~\ref{fig15}(a) is very close to this critical state. If $R<R_s$, the shape Fig.~\ref{fig15}(a) will change. But what kind of structure will form is still unknown. If we gradually decrease $v$, the central cell's two untouched neighbors will adhere to it when $v<0.92$. Then new phase will form and in which each cell adheres to it's six neighbors, such as Fig.~\ref{fig15}(b). From Fig.~\ref{fig15}(a) to (b) the $D_2$ symmetry remains unchanged. For Fig.~\ref{fig15}(b), the adhesion regions clearly are no longer the straight lines.

When structures are periodic in three directions, the calculation will be difficult because it needs to search the optimal distance in three directions. Our simulation indicates optimal shape possibly has unequal periods on different directions. Moreover, complex formations will appear, in which one periodic unit will contain many cells. It leads the problem to be more intricate. Therefor, the study of complex stacking system is still in challenging.

\section{Conclusions} \label{Conclusions}

We have studied the cell stacking system due to adhesion in
2D case. Compared with the random foam model, our model is more fit the reality of biotissues.
We derive the analytical results for three kinds of periodic formations without volume constraint. The corresponding mechanical
parameters, such as the Young¡¯s modulus, bulk modulus and failure strength, are obtained. We provide a potential method to connect the microstructure with the macro-mechanical parameters of biotissues.
We also find that the increase of cellular volume will induce the $D_2$ symmetrical bead-like shape changing to squiggly shapes. These waved shapes imply there may be stable helical formations in 3D case. Our simulation reveals that there are plentiful degenerate energy shapes when many cells adhere together. The adhesion systems are flexible to transfer between these degenerate states. For periodic bead-like shape and foursquare shape with fixed volume, the lowest energy formations are nearby the critical states. Especially, there is a globally optimal solution for periodic bead-like shapes. We think the above results are helpful to understand the physical mechanism for the formation of biotissues. But it needs a more complicate investigation in 3D case, which will be our future work.

\begin{acknowledgements}
We would like to thank professor Ken Brakke for his kind help on the use of the \emph{Surface Evolver}. This work is supported by the National Natural Science Foundation of
China Grants 11304383, 11304241 and 11374237.
\end{acknowledgements}

\end{document}